\documentclass[10pt]{iopart}

\usepackage{booktabs}
\usepackage{xy}

\usepackage{amsmath}
\usepackage{amsthm}
\usepackage{amssymb}
\input xy
\xyoption{all}

\usepackage{hyperref}  

\numberwithin{equation}{section}
\newtheorem{theorem}{Theorem}
\newtheorem{lemma}{Lemma}
\newtheorem{definition}{Definition}

\begin{document}

\title[A reduced subduction graph and higher multiplicity in simmetric groups]{A reduced subduction graph and higher multiplicity in $S_n$ transformation coefficients}

\author{Vincenzo Chilla}

\address{Dipartimento di Fisica ``Enrico Fermi'', Universit\`a di Pisa and Sezione INFN  - Largo Bruno Pontecorvo 3, 56127 Pisa, Italy}
\ead{chilla@df.unipi.it}

\begin{abstract}
Transformation coefficients between {\it standard} bases for irreducible representations of the symmetric group $S_n$ and {\it split} bases adapted to the $S_{n_1} \times S_{n_2} \subset S_n$ subgroup ($n_1 +n_2 = n$) are considered. We first provide a \emph{selection rule} and an \emph{identity rule} for the subduction coefficients which allow to decrease the number of unknowns and equations arising from the linear method by Pan and Chen. Then, using the {\it reduced subduction graph} approach, we may look at higher multiplicity instances. As a significant example, an orthonormalized solution for the first multiplicity-three case, which occurs in the decomposition of the irreducible representation $[4,3,2,1]$ of $S_{10}$ into $[3,2,1] \otimes [3,1]$ of $S_6 \times S_4$, is presented and discussed. \\
\phantom{a} 
\end{abstract}

\section{Introduction}
Symmetric group transformation coefficients, which define various basis changing between $S_n$ representations, are very useful both by itself and in relation to the unitary group representation theory via the Schur-Weyl duality~\cite{Kramer, Haase}. The full comprehension of the multiplicity separation problem is a key outstanding issue. Probably, the main question remains whether there are combinatoric labels which provides a canonical separation of the multiplicity.   
\par  
{\it Subduction coefficients} (SDCs) represent the entries of the transformation matrix between the {\it standard Young-Yamanouchi basis} for an irreducible representation (irrep) of $S_n$ and the {\it split basis} adapted to the subgroup $S_{n_1} \times S_{n_2} \subset S_n $, with $n_1 + n_2 = n$~\cite{McAven}. Such coefficients were introduced since 1953 by Elliot {\it et al}~\cite{Elliot} but, although their calculation have been undertaken for many time~\cite{Kaplan, Kramer1, Rao, Chen1, McAven1, McAven2}, there is still a need for more efficient approaches. 
\par 
Pan and Chen~\cite{Chen} have presented the {\it linear equation method} that is particularly useful since it provides $q$-dependent algebraic solutions for Hecke algebra $H_n(q)$, a quantum deformation of the group algebra $\mathbb{C} S_n$. In~\cite{Chilla} we have given an improved version of such a method which uses the concept of {\it subduction graph} to select a minimal set of linear equations solving the subduction problem for symmetric groups in a systematic manner. In this paper we look for a more insight into the structure of the solution for such a system, reducing the number of unknown SDCs and the number of needed equations. It allows to increase the dimension of the involved  irreps and thus to find solutions for higher multiplicity cases.     
\par 
The layout of the paper is as follows. In the next section, we review some background and we refer the reader to~\cite{Chilla}, and references therein, for definitions, notations and for more details on the subduction graph method. In section 3, we analyze the structure of the subduction space and we prove two theorems which are useful to reduce the number of unknowns and, consequently, the number of equations for the subduction problem. That is fundamental for an optimazed approach to very high dimension decompositions. According to McAven {\it et al}~\cite[pg 8372]{McAven}, we think that ``the next steps in a search for a combinatorial recipe for a multiplicity separation could be to look at other multiplicity two cases and the first multiplicity three case''. Therefore in section 4, we present our determination for the significant first multiplicity-three examples in the reduction $S_{10} \downarrow S_6 \times S_4$. Finally, in section 5, our results are summarized.
\section{Subduction coefficients and graphs} 
The irreducible representations (irreps) of the symmetric group $S_n$ may be labelled by partitions $[\lambda]$ of $n$, i.e. sequences $[\lambda_1, \lambda_2, \ldots, \lambda_h]$ of positive integers such that $\sum_{i=1}^h \lambda_i = n$ and the $\lambda_i$ are weakly decreasing. A partition $[\lambda]$ is usually represented by a Ferrers diagram (or Young diagram) obtained from a left-justified array with $\lambda_j$ boxes on the $j$th row and with the $k$th row below the $(k-1)$th row. Standard Young tableaux are generated by filling the Ferrers diagram with the numbers $1, \ldots, n$ in such a way that each number appears exactly once and the numbers are strictly increasing along the rows and down the columns. An orthonormal basis vector of an irrep associated to the partition $[\lambda]$ may be labelled by a standard Young tableau. Such a basis corresponds to the Gelfand-Tzetlin chain $ S_1 \subset S_2 \subset \ldots \subset S_n$ and is usually called the \emph{standard basis} of $[\lambda]$. We denote this basis by \emph{$S_n$-basis}~\cite{McAven}.
\par
\emph{Split basis}~\cite{McAven} represents an  alternative orthonormal basis for $[\lambda]$. By definition, it breaks $[\lambda]$ (which is, in general, a reducible representation of the direct product subgroup $S_{n_1} \times S_{n_2}$, with $n_1 + n_2 =n$) in a block-diagonal form: 
\begin{equation}
[\lambda]=\bigoplus_{\lambda_1,\lambda_2} \{\lambda; \lambda_1, \lambda_2\} \ [\lambda_1]\otimes[\lambda_2],
\end{equation}
where [$\lambda_1]$ and $[\lambda_2]$ are irreps of $S_{n_1}$ and $S_{n_2}$ respectively. $\{\lambda; \lambda_1, \lambda_2\}$ give the \emph{multiplicity}, i.e. the number of times that the irrep $[\lambda_1] \otimes [\lambda_2]$ of  $S_{n_1} \times S_{n_2}$ appears in the decomposition of $[\lambda]$. 
The entries of the matrix transforming between split and standard basis are the \emph{subduction coefficients} (SDCs). Let $[\lambda_1] \otimes [\lambda_2]$ be a \emph{fixed} irrep of $S_{n_1}  \times S_{n_2}$ in $[\lambda] \downarrow S_{n_1} \times S_{n_2}$ and $| \lambda_1, \lambda_2 ; m_1, m_2 \rangle_{\eta}$  a generic vector of the split basis (where $m_1$ and $m_2$ are standard Young tableaux with Ferrers diagram $\lambda_1$ and $\lambda_2$ respectively,  and $\eta$ is the multiplicity label). We may expand such vectors in terms of the standard basis vectors $|\lambda ; m \rangle $ of $[\lambda]$:
\begin{equation}
| \lambda_1, \lambda_2 ; m_1, m_2 \rangle_{\eta} = \sum_m \ |\lambda ; m \rangle \langle \lambda ; m | \lambda_1, \lambda_2 ; m_1, m_2 \rangle_{\eta}. 
\end{equation}
Thus $\langle \lambda ; m | \lambda_1, \lambda_2 ; m_1, m_2 \rangle_{\eta}$ represent the SDCs of $[\lambda] \downarrow [\lambda_1] \otimes [\lambda_2]$ with given multiplicity label $\eta$ and satisfy the following unitary conditions~\cite{Chenbook2}:
\begin{equation}
\sum_m \ \langle \lambda ; m | \lambda_1, \lambda_2 ; m_1, m_2 \rangle_{\eta} \ \langle \lambda ; m | \lambda_1, \lambda'_2 ; m_1, m'_2 \rangle_{\eta'} = \delta_{\lambda_2 \lambda'_2} \delta_{m_2 m'_2} \delta_{\eta \eta'}
\label{orton1}
\end{equation}
\begin{equation}
\sum_{\lambda_2 m_2 \eta} \ \langle \lambda ; m | \lambda_1, \lambda_2 ; m_1, m_2 \rangle_{\eta} \ \langle \lambda ; m' | \lambda_1, \lambda_2 ; m_1, m_2 \rangle_{\eta} = \delta_{m m'}.
\label{orton2}
\end{equation}
\par
Given a standard Young tableau $m$, we define the action $g_i(m)$ of a generator $g_i$ for $S_n$ (elementary transposition) in the following way: if the tableau obtained from $m$ interchanging the box with $i$ and the box with $i+1$ (keeping the other elements fixed) is another standard Young tableau $m'$, we set $g_i(m)=m'$; else $g_i(m)=m$. Furthermore, we define 
\begin{equation}
g_i(m_{12}) = \left\{
\begin{array}{cc}
(g_i(m_1),m_2) & \text{if $i < n_1$} \\
(m_1, g_i(m_2))& \text{if $i > n_1$}
\end{array}
\right.
\end{equation}
where $m_{12}=(m_1, m_2)$ is a pair of Standard Young tableaux with $n_1$ and $n_2$ boxes respectively, with $m_1$ filled by integers from $1$ to $n_1$ and $m_2$ from $n_1+1$ to $n_1 + n_2$
\par 
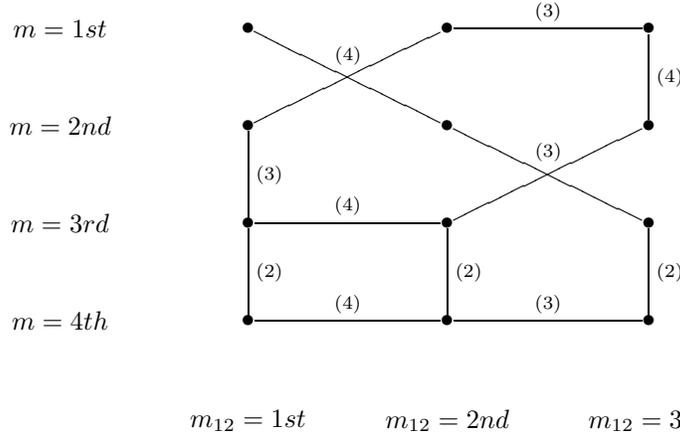
\begin{figure}
$\xymatrix{
m=1st & *{\bullet} \ar@{-}[rd]^(0.45){(4)} & *{\bullet} \ar@{-}[ld] \ar@{-}[r]^{(3)} & *{\bullet}  \ar@{-}[d]^{(4)} \\    
 m=2nd &   *{  \bullet} \ar@{-}[d]^{(3)} & *{\bullet} \ar@{-}[rd]^(0.45){(3)} & 
*{\bullet} \ar@{-}[ld]  \\
m=3rd & *{ \bullet} \ar@{-}[r]^{(4)} \ar@{-}[d]^{(2)}  & *{\bullet} \ar@{-}[d]
^{(2)} & *{\bullet} \ar@{-}[d]^{(2)} \\
m=4th & *{ \bullet} \ar@{-}[r]^{(4)} & *{\bullet}\ar@{-}[r]^{(3)} & *{\bullet} \\
 & m_{12}=1st & m_{12}=2nd & m_{12}=3rd}$
\caption{\small{Subduction graph relative to $([4,1]; [1], [3,1])$.}}
\label{subdgraf}
\end{figure}
Two standard Young tableaux $m_1$ and $m_2$ with the same Ferrers diagram are \emph{$i$-coupled} if $m_1 = m_2$ or if $m_1=g_i(m_2)$. In an analogous way, we say that the pairs of standard Young tableaux $m_{12}=(m_1,m_2)$ and $m_{34}=(m_3,m_4)$ are \emph{$i$-coupled} if $m_{12} = m_{34}$ or if $g_i(m_{12}) = m_{34}$. \par
Each ordered sequence of three standard Young tableaux $(m; m_1, m_2)$ with Ferrers diagrams $\lambda$, $\lambda_1$ and $\lambda_2$ respectively and filled as previously described is called \emph{node} and it is denoted by $\langle m; m_1, m_2 \rangle$, or simply $\langle m; m_{12} \rangle$. Moreover, we call \emph{subduction grid} (or simply \emph{grid}) the set of all nodes of $(\lambda; \lambda_1, \lambda_2)$. Thus, in analogy with the case of standard Young tableaux, we may define the action of $g_i$ on a node $\nu = \langle m; m_{12} \rangle $ as
\begin{equation}
g_i(\nu)= \langle g_i(m); g_i(m_{12}) \rangle.    
\end{equation}
Two nodes $\nu_1$ and $\nu_2$ are \emph{i-coupled} if $\nu_1 = \nu_2$ or if $\nu_1 = g_i(\nu_2)$. 
For $i$ fixed, $i$-coupling is an equivalence relation on the grid. 
\par 
Therefore, the $i$-coupling relation induces a partition on the grid which is called \emph{i-layer} and we may give a graphical representation of it. Each node is represented  in such a way that it has coordinates obtained by the ordering number of the relative standard Young tableau (for example the lexicographic ordering~\cite{Chenbook}). Furthermore, because each equivalence class is composed at most by two distinct nodes, we may draw them as joined by an edge with a label for $i$.  
\par
The \emph{overlap} of all $i$-layers (i.e. the graph obtained by identification of the corresponding  nodes) is called \emph{subduction graph} relative to  $(\lambda; \lambda_1, \lambda_2)$. More simply, two distinct nodes $\nu$ and $\nu'$ of the grid are connected by an edge with the label $(i)$ of the subduction graph if $\nu=g_i(\nu')$ for some $i$ (notice that if $\nu$ and $\nu'$ are $i$-coupled and $j$-coupled, then $i=j$). 
In figure~\ref{subdgraf}, the graphical representation of the subduction graph for $([4,1]; [1], [3,1])$, obtained from the overlap of the $2$-layer, the $3$-layer and the $4$-layer, is shown as an example. Note that the $1$-layer is not defined for $([4,1]; [1], [3,1])$ due to $[\lambda_1]=[1]$ (see also~\cite{Chilla}). 
\par
We observe that each node $\langle m; m_{12} \rangle$ relative to $(\lambda; \lambda_1, \lambda_2)$ naturally corrisponds to the SDC $\langle \lambda ; m| \lambda_1, \lambda_2 ; m_1,m_2 \rangle$ given by the inner product between the standard basis vector $|\lambda; m\rangle$ for the irrep $[\lambda]$ of $S_n$ and the split basis vector $|\lambda_1, \lambda_2 ; m_1, m_2 \rangle$. Furthermore, subduction graph provides a very practical way to choose a minimal set of subduction equations to find the required transformation coefficients~\cite{Chilla}. 
\section{Selection and identity rules: reduced subduction graph}
\subsection{Crossing and bridge pairs of standard Young tableaux}
Let $\lambda$ be a Young diagram relative to a partition of $n$ and $(m,m')$ a pair of standard Young tableaux with the same diagram $\lambda$. Furthermore, we denote by $d_k(m)$ the usual \emph{axial distance} between the numbers $k$ and $k+1$ in the tableau $m$. \\
If $m \neq m'$, we name \emph{cut} the minimum $i \in \{1, \ldots, n-1\}$ such that $d_i(m) \neq d_i(m')$. We give the following useful definitions:
\begin{definition}
We say that $(m,m')$ is a \emph{crossing pair} of standard Young tableaux if there exists $i \in \{1, \ldots, n-1 \}$ such that one of the following cases is verified: 
\begin{enumerate}
\item $d_i(m) \neq d_i(m')$, $g_i(m)\neq m$ and $ g_i(m') \neq m'$;
\item $d_i(m) \neq d_i(m')$, $g_i(m) = m$ and $g_i(m') = m'$.
\end{enumerate}
We call \emph{separation} for $(m,m')$ the minimum $i$ where one of the previous cases occurs.  
\label{cros}
\end{definition}
\begin{definition}
We say that $(m,m')$ is a \emph{bridge pair} of standard Young tableaux if it is \emph{not} a crossing pair, i.e.  
for all $i \in \{1, \ldots, n-1 \}$ one of the following cases is verified: 
\begin{enumerate}
\item $d_i(m)=d_i(m')$;
\item $g_i(m)=m$ and $ g_i(m') \neq m'$;
\item $ g_i(m) \neq m$ and $g_i(m')=m'$.
\end{enumerate}
\label{brid} 
\end{definition}
\begin{lemma}
Let $(m, m')$ be a bridge pair with $m \neq m'$ and let $\bar{i}$ be the relative cut. Let us consider the application defined by
\begin{equation}
g_{\bar{i}}(m,m')= (g_{\bar{i}}(m),g_{\bar{i}}(m')).
\label{bridge}
\end{equation}
Then, by iteratively applying (\ref{bridge}), we always obtain a crossing pair.
\label{ponti}
\end{lemma}
\begin{proof}
We first observe that, after one application of $g_{\bar{i}}$ on $(m,m')$, we have the following situation
\begin{equation}
\left\{ 
\begin{array}{cc}
d_j(g_{\bar{i}}(m)) = d_j(g_{\bar{i}}(m')) & \text{if $j \notin \{\bar{i}-1,\bar{i},\bar{i}+1\}$}  \\
d_j(g_{\bar{i}}(m)) = d_{j}(g_{\bar{i}}(m')) + d_{j+1}(g_{\bar{i}}(m')) & \text{if $j=\bar{i}-1$ }  \\
d_j(g_{\bar{i}}(m)) = - d_j(g_{\bar{i}}(m')) & \text{if $j=\bar{i}$ } \\
d_j(g_{\bar{i}}(m)) = d_{j-1}(g_{\bar{i}}(m')) + d_{j}(g_{\bar{i}}(m')) & \text{if $j=\bar{i}+1$ }   
\end{array}
\right.
\label{dsep} 
\end{equation}
thus $g_{\bar{i}}(m,m')$ has cut in $\bar{i}-1$ because obviously $d_{\bar{i}}(g_{\bar{i}}(m')) \neq 0$. \\
Then, at each step of the iteration of (\ref{bridge}), two cases may occur:
\begin{enumerate}
\item $g_{\bar{i}}(m,m')$ is a crossing pair and we have the assertion. 
\item $g_{\bar{i}}(m,m')$ is a bridge pair with cut in $\bar{i}-1$. 
\end{enumerate}
If case $(i)$ never occurs, after $\bar{i}-1$ iterations we should reach a bridge pair $(\tilde{m},\tilde{m}')$ with cut $i=1$. But $(\tilde{m},\tilde{m}')$ always is a crossing pair because $g_1(\tilde{m})=\tilde{m}$ and $g_1(\tilde{m}')=\tilde{m}'$ for each standard Young tableaux $\tilde{m}$ and $\tilde{m}'$.  
\end{proof}
\subsection{Islands}
Let $m$, $m_1$ and $m_2$ be three standard Young tableaux with $n$, $n_1$ and $n_2$ boxes such that $n_1 + n_2 = n$ and shapes $\lambda$, $\lambda_1$ and $\lambda_2$, respectively. 
Denoted by $m^{(n_1)}$ the standard Young tableau obtained from $m$ by removing the boxes with numbers $n_1+1, \ldots, n$, we say that $m$ and $m_1$ are \emph{compatible} if $m_1=m^{(n_1)}$.
 The number of standard Young tableaux which are compatible with $m_1$ is equal to the number of standard skew-tableaux~\cite{Fulton} of shape $\lambda/\lambda_1$ filled with the numbers $n_1 +1, ..., n$. We denote it by $f^{\lambda/\lambda_1}$. 
\par 
Denoted by $G$ the grid relative to $(\lambda; \lambda_1, \lambda_2)$, we give the following
\begin{definition}
Fixed the standard tableau $\mu$ with Young diagram $\lambda_1$ and varying $m$ and $m_2$, with fixed Young diagrams $\lambda$ and $\lambda_2$ respectively, the subset of $G$ given by 
\begin{equation}
I_{\mu}(G) = \{\langle m; \mu, m_2 \rangle \ \in G \ | \ \ m \ \text {is compatible with} \ \mu \}
\end{equation}
is named \emph{$\mu$-island}of $G$.
\end{definition}
We refer to the $\mu$-island simply saying {\it island} if it is not necessary to make an explicit reference to $\mu$. Of course, the number of islands of $G$ is given by the number of standard Young tableaux with diagram $\lambda_1$, i.e. $f^{\lambda_1}$.   
\begin{lemma}
Let $\langle m; m_{12} \rangle \in G $ be a node such that $(m^{(n_1)},m_1)$ is a crossing pair. Then the corresponding SDC, $\langle \lambda; m | \lambda_1, \lambda_2 ; m_1, m_2 \rangle$, vanishes.  
\label{incroci}
\end{lemma}
\begin{proof}
Let $i \in \{1, \ldots, n_1 -1\}$ be the separation of $(m^{(n_1)},m_1)$. From definition \ref{cros}, we need to destinguish the following situations:
\begin{itemize}
\item $g_i(m) \neq m $ and $g_i(m_1) \neq m_1 $ (or, equivalently, $d_i(m)\neq \pm 1$ and $d_i(m_1) \neq \pm 1$). \\
The action of the generator $g_i$ on the standard base vector $|\lambda; m \rangle$ is given by~\cite{Chen}
\begin{equation}
g_i |\lambda; m \rangle = \frac{1}{d_i(m)} \ |\lambda; m \rangle \ + \  \beta^{(i)}_m \ |\lambda; g_i(m) \rangle
\label{action1}
\end{equation}
where 
\begin{equation}
\beta^{(i)}_m = \sqrt{1-\frac{1}{d_i^2(m)}}.
\end{equation}
In an analogous way, the action on the split base vector $ |\lambda_1, \lambda_2;m_1,m_2 \rangle$ is
\begin{equation}
g_i |\lambda_1, \lambda_2 ; m_1, m_2 \rangle = \frac{1}{d_i(m_1)} \ |\lambda_1,\lambda_2; m_1, m_2 \rangle \ + \ \beta^{(i)}_{m_1} \ |\lambda_1,\lambda_2; g_i(m_1), m_2 \rangle
\label{action2}
\end{equation}
From (\ref{action1}) and (\ref{action2}), using $g_i^2 = 1$ and $g_i={g_i}^{\dagger}$, we get
\begin{equation}
\left(1-\frac{1}{d_i(m) d_i(m_1)}\right) \ \langle \lambda;m|\lambda_1, \lambda_2 ;m_1, m_2 \rangle \ - \ \frac{\beta^{(i)}_{m_1}}{d_i(m)} \  \langle \lambda; m |\lambda_1, \lambda_2 ; g_i(m_1), m_2 \rangle \ + 
\nonumber
\end{equation}
\begin{equation}
- \ \frac{\beta^{(i)}_{m}}{d_i(m_1)} \  \langle \lambda; g_i(m)|\lambda_1,\lambda_2 ; m_1, m_2 \rangle \ - \ \beta^{(i)}_m \beta^{(i)}_{m_1} \ \langle\lambda; g_i(m)|\lambda_1, \lambda_2; g_i(m_1), m_2 \rangle  = 0.
\label{isol}
\end{equation}
Writing (\ref{isol}) also for $\langle\lambda; m |\lambda_1, \lambda_2;g_i(m_1), m_2  \rangle$, $\langle \lambda;g_i(m)|\lambda_1, \lambda_2; m_1, m_2  \rangle$ and $\langle\lambda; g_i(m)|\lambda_1, \lambda_2 ; g_i(m_1), m_2  \rangle$ and using $d_i(g_i(m))=-d_i(m)$ and $d_i(g_i(m_{12}))=-d_i(m_{12})$, we obtain the homogeneous linear system described by the matrix 
\begin{equation}
\left(
\begin{array}{cccc}
1-\frac{1}{d_i(m) d_i(m_1)} & -\frac{\beta^{(i)}_{m_1}}{d_i(m)} & -\frac{\beta^{(i)}_{m}}{d_i(m_1)} & -\beta^{(i)}_m \beta^{(i)}_{m_1} \\
 -\frac{\beta^{(i)}_{m_1}}{d_i(m)} & 1+\frac{1}{d_i(m) d_i(m_1)} & -\beta^{(i)}_m \beta^{(i)}_{m_1} & \frac{\beta^{(i)}_{m}}{d_i(m_1)}\\
-\frac{\beta^{(i)}_{m}}{d_i(m_1)} & -\beta^{(i)}_m \beta^{(i)}_{m_1} & 1+\frac{1}{d_i(m) d_i(m_1)}  & \frac{\beta^{(i)}_{m_1}}{d_i(m)} \\
 -\beta^{(i)}_m \beta^{(i)}_{m_1} & \frac{\beta^{(i)}_{m}}{d_i(m_1)} & \frac{\beta^{(i)}_{m_1}}{d_i(m)} & 1-\frac{1}{d_i(m) d_i(m_1)}
\end{array}
\right).
\label{matr}
\end{equation}
Thus $\langle\lambda; m|\lambda_1, \lambda_2; m_1, m_2 \rangle$,  $\langle\lambda; m |\lambda_1, \lambda_2;g_i(m_1), m_2  \rangle$, $\langle \lambda;g_i(m)|\lambda_1, \lambda_2; m_1, m_2  \rangle$ and $\langle\lambda; g_i(m)|\lambda_1, \lambda_2 ; g_i(m_1), m_2  \rangle$ are the coordinates of a kernel vector for the matrix (\ref{matr}). It is easy to see that (\ref{matr}) has rank $3$.
\begin{enumerate}
\item If $d_i(m) \neq - d_i(m_{1})$;\\
the kernel space for (\ref{matr}) is generated by the vector
\begin{equation}
\left(
\begin{array}{c}
1 \\
 \frac{d_i(m) d_i(m_1)(\beta^{(i)}_m + \beta^{(i)}_{m_1})}{d_i(m) + d_i(m_1)} \\
 \frac{d_i(m) d_i(m_1)(\beta^{(i)}_m + \beta^{(i)}_{m_1})}{d_i(m) + d_i(m_1)} \\
- 1
\end{array}
\right),
\end{equation}
which implies 
\begin{equation}
\langle \lambda;m |\lambda_1,\lambda_2 ; m_1, m_2 \rangle = - \langle \lambda; g_i(m) |\lambda_1, \lambda_2; g_i(m_1), m_2 \rangle
\label{cond1}
\end{equation} 
 and 
\begin{equation}
 \langle \lambda; g_i(m)|\lambda_1, \lambda_2; m_1, m_2 \rangle = \langle \lambda;m |\lambda_1,\lambda_2; g_i(m_1), m_2 \rangle.
\label{cond2}
\end{equation}
Because $d_i(g_i(m)) = -d_i(m)$, $d_i(m) \neq - d_i(m_1) \Rightarrow d_i(g_i(m)) \neq d_i(m_1)$ and $d_i(m) \neq d_i(m_1) \Rightarrow d_i(g_i(m)) \neq - d_i(m_1)$. Therefore relation (\ref{cond2}), written for $\langle \lambda; g_i(m) | \lambda_1 , \lambda_2 ;m_1, m_2 \rangle$, yelds (remember that $g_i^2 = 1$)
 \begin{equation}
 \langle \lambda ; m |\lambda_1, \lambda_2 ; m_1, m_2 \rangle = \langle\lambda; g_i(m)| \lambda_1, \lambda_2 ; g_i(m_1), m_2 \rangle.
 \label{cond5} 
 \end{equation}
 From (\ref{cond1}) and (\ref{cond5}), we get 
\begin{equation}
\langle\lambda; m|\lambda_1, \lambda_2;m_1, m_2 \rangle = 0.
\end{equation}
\item If $d_i(m) = -d_i(m_1)$; \\
the kernel space for (\ref{matr}) is generated by the vector
\begin{equation}
\left(
\begin{array}{c}
0 \\
1 \\
1 \\
0
\end{array}
\right)
\end{equation}
which directly implies
\begin{equation}
\langle \lambda;g_i(m)| \lambda_1, \lambda_2 ; m_1, m_2 \rangle = \langle\lambda; m|\lambda_1, \lambda_2; g_i(m_1), m_2 \rangle
\label{cond3}
\end{equation}
and
\begin{equation}
\langle \lambda; m|\lambda_1, \lambda_2 ;m_1, m_2 \rangle = \langle \lambda; g_i(m)| \lambda_1, \lambda_2; g_i(m_1), m_2 \rangle = 0.
\label{cond2bis}
\end{equation} 
\end{enumerate}
\item $g_i(m) = m$ and $g_i(m_1) = m_1$  (or, equivalently, $|d_i(m)|=1$ and $|d_i(m_1)|=1$). \\
The action of the generator $g_i$ on the standard base vector $|\lambda; m \rangle$ is given by
\begin{equation}
g_i |\lambda; m \rangle = \pm |\lambda; m \rangle
\end{equation}
and the action on the split base vector $ |\lambda_1, \lambda_2;m_1,m_2 \rangle$ is
\begin{equation}
g_i |\lambda_1, \lambda_2 ; m_1, m_2 \rangle = \mp|\lambda_1,\lambda_2; m_1, m_2 \rangle. 
\end{equation}
Thus (because $g_i^2=1$ and $g_i={g_i}^{\dagger}$)
\begin{equation}
\langle \lambda;m|\lambda_1, \lambda_2;m_1,m_2 \rangle = - \langle \lambda;m|\lambda_1, \lambda_2;m_1,m_2 \rangle
\end{equation}
from which
\begin{equation}
\langle \lambda;m|\lambda_1, \lambda_2;m_1,m_2 \rangle = 0.
\end{equation}
\end{itemize}
\end{proof}
\begin{lemma}
Let $\langle m; m_{12} \rangle \in G $ be a node such that $(m^{(n_1)},m_1)$ is a bridge pair and $m^{(n_1)}\neq m_1$. Then the corresponding SDC, $\langle \lambda; m | \lambda_1, \lambda_2 ; m_1, m_2 \rangle$, vanishes. 
\label{ponti1}
\end{lemma}
\begin{proof}
Let $i \in \{1, \ldots, n_1 -1\}$ be the cut of $(m^{(n_1)},m_1)$. For semplicity, let us suppose $g_i(m) = m $ and $g_i(m_1)\neq m_1 $. \\
The action of the generator $g_i$ on the standard base vector $|\lambda; m \rangle$ is given by
\begin{equation}
g_i |\lambda; m \rangle = \pm |\lambda; m \rangle 
\label{action3}
\end{equation}
and the action on the split base vector $ |\lambda_1, \lambda_2;m_1,m_2 \rangle$ is
\begin{equation}
g_i |\lambda_1, \lambda_2 ; m_1, m_2 \rangle = \frac{1}{d_i(m_1)} \ |\lambda_1,\lambda_2; m_1, m_2 \rangle \ + \ \beta^{(i)}_{m_1} \ |\lambda_1,\lambda_2; g_i(m_1), m_2 \rangle
\label{action4}
\end{equation} 
Because $d_i(m_1)\neq \pm 1$, (\ref{action3}) and (\ref{action4}) imply 
\begin{equation}
\langle \lambda;m|\lambda_1, \lambda_2;m_1,m_2 \rangle = b'_{i,m_1} \langle \lambda;m|\lambda_1, \lambda_2;g_i(m_1),m_2 \rangle
\label{it1}
\end{equation}
with $b'_{i,m_1}$ a suitable numerical factor.\\ 
In an analogous way, the case $g_i(m) \neq m$ and $g_i(m_1)=m_1$ provides
\begin{equation}
\langle \lambda;m|\lambda_1, \lambda_2;m_1,m_2 \rangle = b''_{i,m} \langle \lambda;g_i(m)|\lambda_1, \lambda_2;m_1,m_2 \rangle
\label{it2}
\end{equation}
with $b''_{i,m}$ another suitable numerical factor. \\
From lemma \ref{ponti}, by iterating the previous derivation, we may write
\begin{equation}
\langle \lambda;m|\lambda_1, \lambda_2;m_1,m_2 \rangle = b \ \langle \lambda; \bar{m} |\lambda_1, \lambda_2;\bar{m}_1,m_2 \rangle 
\end{equation}
with $(\bar{m}, \bar{m}_1)$ a crossing pair and $b$ a total numerical factor . But, from lemma \ref{incroci},
\begin{equation}
\langle \lambda; \bar{m} |\lambda_1, \lambda_2;\bar{m}_1,m_2 \rangle = 0,
\end{equation}
thus 
\begin{equation}
 \langle \lambda;m|\lambda_1, \lambda_2;m_1,m_2 \rangle = 0.
\end{equation}
\end{proof}
\par 
It is now possible to give the following 
\begin{theorem}[Selection Rule]
Let $\langle m; m_{12} \rangle$ be a node of $G$ which does not belong to any island of $G$. Then the corresponding SDC, $\langle \lambda; m | \lambda_1, \lambda_2 ; m_1, m_2 \rangle$, vanishes.
\end{theorem}
\begin{proof}
Because $\langle m; m_{12} \rangle$ does not belong to any island, we have $m^{(n_1)} \neq m_1$. If $(m^{(n_i)},m_1)$ is a crossing pair, we have the assertion by lemma \ref{incroci}. If $(m^{(n_i)},m_1)$ is not a crossing pair (i.e. it is a bridge pair), we use lemma \ref{ponti1} and we have the proof. 
\end{proof}
The previous theorem allows us to say that only $f^{\lambda_1} f^{\lambda_2} f^{\lambda/\lambda_1}$ SDCs may not vanishes. It provides a selection rule for the subduction coefficients which is based on the Littlewood-Richardson rule. Furthermore, we observe that, in our graph approach, it is analogous to the \emph{block-selective rule} given in~\cite{McAven1} and~\cite{McAven2} . Therefore we may in somehow associate our definition of island to the concept of ``block'' given by McAven and Butler.   
\par We now show another simple proposition that goes further on reducing the number of unknown SDCs.
\begin{theorem}[Identity Rule]
All islands of $G$ have the same corresponding SDCs, i.e. 
\begin{equation}
\langle\lambda; m|\lambda_1, \lambda_2; m_1, m_2 \rangle = \langle \lambda; g_i(m)| \lambda_1,\lambda_2; g_i(m_1), m_2 \rangle
\nonumber
\end{equation}
for all $i\in \{1, \ldots, n_1 -1 \}$.
\end{theorem}
\begin{proof}
Suppose $\langle m;m_{12} \rangle$ belongs to the $m_1$-island. Thus $m$ is compatible with $m_1$ and we have $d_i(m)=d_i(m_1)=d_i$ for all $i\in \{1, \ldots, n_1-1\}$. We again distinguish two cases: 
\begin{itemize}
\item $d_i \neq \pm 1$.\\  It is straightforward that (\ref{matr}) is a rank $2$ matrix and the kernel space is generated by the vectors
\begin{equation}
\left(
\begin{array}{c}
1 \\
0 \\
0 \\
1
\end{array}
\right), \ \ \ 
\left(
\begin{array}{c}
\frac{2}{\sqrt{d_i^2 -1 }} \\
1 \\
1 \\
0
\end{array}
\right)
\end{equation}
therefore we have
\begin{equation}
\langle\lambda;  g_i(m)|\lambda_1,\lambda_2;m_1,m_2 \rangle = \langle\lambda; m| \lambda_1,\lambda_2;g_i(m_1), m_2 \rangle
\label{cond2tris}
\end{equation}
and
\begin{equation}
\langle \lambda;m|\lambda_1,\lambda_2; m_1,m_2 \rangle = \langle \lambda;g_i(m)|\lambda_1,\lambda_2; g_i(m_1), m_2 \rangle + 
\nonumber
\end{equation}
\begin{equation}
+ \frac{2}{\sqrt{d_i^2 -1 }} \langle\lambda;  g_i(m)| \lambda_1, \lambda_2; m_1, m_2\rangle.
\label{cond4}
\end{equation}
Because $d_i(g_i(m))= - d_i(m)=-d_i(m_1)$, (\ref{cond2bis}) becomes  
\begin{equation}
\langle \lambda; g_i(m)| \lambda_1, \lambda_2; m_1, m_2 \rangle = 0
\end{equation}
and, from (\ref{cond4}),
\begin{equation}
\langle\lambda; m|\lambda_1, \lambda_2; m_1, m_2 \rangle = \langle \lambda; g_i(m)| \lambda_1,\lambda_2; g_i(m_1), m_2 \rangle.
\end{equation}
Thus the $m_1$-island and the $g_i(m_1)$-island have the same corresponding SDCs. 
\item $d_i = \pm 1$.\\ In this case $g_i(m)=m$ and $g_i(m_{12})=m_{12}$, thus both $\langle m; m_{12} \rangle$ and $\langle g_i(m); g_i(m_{12}) \rangle$ trivially belong to the same $m_1$-island and, of course,  
\begin{equation}
\langle \lambda;m| \lambda_1, \lambda_2; m_1,m_2 \rangle = \langle \lambda; g_i(m)| \lambda_1,\lambda_2; g_i(m_1), m_2 \rangle.
\end{equation}
\end{itemize}
So the proof directly follows from the fact that the $m_1$-island can be transformed in to another $m_1'$-island by a suitable composition of $g_i$ transformations ($i \in \{1, \ldots, n_1 -1\}$), the same one which transforms the standard Young tableau $m_1$ to $m_1'$.
\end{proof}
\subsection{Reduced subduction graph}
From the previous theorems, the only SDCs we need to evaluate are the $f^{\lambda_2} f^{\lambda/\lambda_1}$ ones relative to a single island. We have a reduced linear system with $(n_2-1) f^{\lambda_2} f^{\lambda/\lambda_1}$ equations and $f^{\lambda_2} f^{\lambda/\lambda_1}$ unknowns instead of the $(n-2) f^{\lambda} f^{\lambda_1} f^{\lambda_2} $ and $f^{\lambda} f^{\lambda_1} f^{\lambda_2}$ primal ones.
\par  In fact, fixed an island, the relative \emph{reduced subduction graph} is sufficient to provide the required transformation coefficients. Such a graph is obtained by the action of the $g_i$ trasformations, with $i \in \{n_1 +1, \ldots, n-1\}$, on the island nodes only, and thus it allows further on reducing the number of dependent linear equations. On the other hand, the $g_i$ transformations with $i \in \{1, \ldots, n_1 - 1 \}$ link the corresponding nodes of two different islands and thus, by the identity rule, we do not need to consider them.
\section{The first multiplicity-three case}

\begin{table}
\label{tabcoeff}
\begin{center}
\begin{tabular}{ccc|ccc|ccc}
\toprule
\multicolumn{3}{c}{$1^{st}$ multiplicity copy} & \multicolumn{3}{c}{$2^{nd}$ multiplicity copy} & \multicolumn{3}{c}{$3^{rd}$ multiplicity copy} \\
\midrule
$\frac{\sqrt{14}}{64}$ & $-\frac{\sqrt{7}}{16}$ &$\frac{\sqrt{21}}{32} $ &$\frac{5 \sqrt{2}}{64}$ & $- \frac{5}{64}$& $- \frac{5 \sqrt{3}}{64} $&$\frac{3 \sqrt{6}}{64}$ &$\frac{3 \sqrt{3}}{64}$ &$\frac{3}{64}$ \\
$-\frac{5 \sqrt{42}}{192}$ &$-\frac{\sqrt{21}}{48}$ &$\frac{3 \sqrt{7}}{32}$ &$- \frac{5 \sqrt{6}}{192}$ &$\frac{5 \sqrt{3}}{192}$ & $- \frac{15}{64}$&$\frac{5 \sqrt{2}}{64}$ &$\frac{13}{64}$ &$\frac{3 \sqrt{3}}{64}$ \\
$\frac{\sqrt{42}}{64}$&$\frac{\sqrt{21}}{32}$ & $0$ &$\frac{5 \sqrt{6}}{64}$ &$- \frac{5 \sqrt{3}}{64}$ & $-\frac{15}{64}$ &$\frac{9 \sqrt{2}}{64}$ &$\frac{3}{64}$ &$\frac{5 \sqrt{3}}{64}$ \\
$\frac{\sqrt{70}}{64}$&$\frac{\sqrt{35}}{32}$ &$0$ &$-\frac{3\sqrt{10}}{64}$ &$\frac{3 \sqrt{5}}{64}$ &$- \frac{5 \sqrt{15}}{64}$ &$\frac{\sqrt{30}}{192}$ &$\frac{11 \sqrt{15}}{192}$ &$\frac{5 \sqrt{5}}{64}$ \\
$-\frac{5 \sqrt{70}}{192}$&$\frac{\sqrt{35}}{24}$ &$\frac{\sqrt{105}}{96}$ & $- \frac{5 \sqrt{10}}{192}$&$- \frac{19\sqrt{5}}{192}$ &$- \frac{7 \sqrt{15}}{192}$ &$\frac{5 \sqrt{30}}{192}$ &$\frac{\sqrt{15}}{192}$ & $\frac{7 \sqrt{5}}{64}$\\
$\frac{\sqrt{210}}{64}$& $0$&$\frac{\sqrt{35}}{32}$ & $- \frac{3 \sqrt{30}}{64}$& $- \frac{5 \sqrt{15}}{64}$&$- \frac{7 \sqrt{5}}{64}$ &$\frac{\sqrt{10}}{64}$ &$\frac{5 \sqrt{5}}{64}$ &$\frac{7 \sqrt{15}}{64}$ \\
$\frac{\sqrt{42}}{64}$&$- \frac{\sqrt{21}}{16}$ &$- \frac{\sqrt{7}}{32}$ &$\frac{5 \sqrt{6}}{64}$ &$- \frac{5 \sqrt{3}}{64}$ &$\frac{5}{64}$ &$\frac{9 \sqrt{2}}{64}$ &$\frac{9}{64}$ &$-\frac{\sqrt{3}}{64}$ \\
$- \frac{5 \sqrt{14}}{64}$&$-\frac{\sqrt{7}}{16}$ & $- \frac{\sqrt{21}}{32}$&$- \frac{5 \sqrt{2}}{64}$ &$\frac{5}{64}$ &$\frac{5 \sqrt{3}}{64}$ &$\frac{5 \sqrt{6}}{64}$ &$\frac{13 \sqrt{3}}{64}$ &$-\frac{3}{64}$ \\
$\frac{\sqrt{70}}{64}$&$0$ & $- \frac{\sqrt{105}}{32}$&$\frac{5 \sqrt{10}}{64}$ &$\frac{3 \sqrt{5}}{64}$ &$- \frac{\sqrt{15}}{64}$ & $\frac{3 \sqrt{30}}{64}$&$-\frac{\sqrt{15}}{64}$ &$\frac{3 \sqrt{5}}{64}$ \\
$- \frac{\sqrt{2}}{64}$&$\frac{1}{8}$ &$- \frac{7 \sqrt{3}}{32}$ & $\frac{\sqrt{14}}{64}$&$\frac{7 \sqrt{7}}{64}$ &$- \frac{\sqrt{21}}{64}$ &$-\frac{\sqrt{42}}{64}$ &$\frac{3 \sqrt{21}}{64}$ &$\frac{3 \sqrt{7}}{64}$\\
$-\frac{7 \sqrt{10}}{64}$&$- \frac{\sqrt{5}}{32}$ &$- \frac{\sqrt{15}}{16}$ &$- \frac{ \sqrt{70}}{64}$ &$\frac{\sqrt{35}}{64}$ &$\frac{\sqrt{105}}{64}$ &$\frac{\sqrt{210}}{64}$ &$-\frac{\sqrt{105}}{64}$ &$\frac{3 \sqrt{35}}{64}$\\
$- \frac{\sqrt{6}}{64}$& $- \frac{7 \sqrt{3}}{32}$& $- \frac{5}{16}$&$\frac{ \sqrt{42}}{64}$ &$- \frac{\sqrt{21}}{64}$ &$\frac{5 \sqrt{7}}{64}$ &$-\frac{3 \sqrt{14}}{64}$ &$\frac{3 \sqrt{7}}{64}$ &$\frac{5 \sqrt{21}}{64}$\\
$\frac{\sqrt{70}}{64}$&$\frac{\sqrt{35}}{32}$ &$0$ &$\frac{5 \sqrt{10}}{64}$ &$-\frac{5 \sqrt{5}}{64}$ &$\frac{3  \sqrt{15}}{64}$ &$\frac{3 \sqrt{30}}{64}$ &$\frac{\sqrt{15}}{64}$ &$-\frac{3 \sqrt{5}}{64}$ \\
$\frac{5 \sqrt{42}}{192}$& $\frac{5 \sqrt{21}}{96}$& $0$& $-\frac{5 \sqrt{6}}{64}$&$\frac{5 \sqrt{3}}{64}$ &$\frac{15}{64}$ &$\frac{5 \sqrt{2}}{192}$ & $\frac{55}{192}$&$-\frac{5 \sqrt{3}}{64}$ \\
$\frac{\sqrt{210}}{64}$& $0$&$\frac{\sqrt{35}}{32}$ &$\frac{5 \sqrt{30}}{64}$ &$\frac{3 \sqrt{15}}{64}$ &$\frac{ \sqrt{5}}{64}$ &$\frac{9\sqrt{10}}{64}$ &$-\frac{3 \sqrt{5}}{64}$&$-\frac{\sqrt{15}}{64}$ \\
$- \frac{\sqrt{6}}{64}$& $\frac{\sqrt{3}}{8}$&$\frac{7}{32}$ &$\frac{\sqrt{42}}{64}$ &$\frac{7 \sqrt{21}}{64}$ &$\frac{\sqrt{7}}{64}$ & $-\frac{3\sqrt{14}}{64}$&$\frac{9 \sqrt{7}}{64}$& $-\frac{\sqrt{21}}{64}$\\
$\frac{7 \sqrt{30}}{192}$&$- \frac{\sqrt{15}}{48}$ & $ \frac{3 \sqrt{5}}{32}$&$-\frac{\sqrt{210}}{64}$ &$\frac{ \sqrt{105}}{64}$ &$\frac{3 \sqrt{35}}{64}$ &$\frac{\sqrt{70}}{192}$ &$-\frac{7 \sqrt{35}}{192}$ &$\frac{\sqrt{105}}{64}$ \\
$- \frac{\sqrt{10}}{64}$& $\frac{\sqrt{5}}{16}$&$\frac{3 \sqrt{15}}{32}$ &$\frac{\sqrt{70}}{64}$ & $-\frac{ \sqrt{35}}{64}$&$\frac{3 \sqrt{105}}{64}$ &$-\frac{\sqrt{210}}{64}$ &$-\frac{\sqrt{105}}{64}$ &$\frac{3 \sqrt{35}}{64}$ \\
$- \frac{35 \sqrt{2}}{192}$& $\frac{7}{24}$&$- \frac{5 \sqrt{3}}{96}$ &$-\frac{5 \sqrt{14}}{192}$ &$-\frac{19 \sqrt{7}}{192}$ &$\frac{5 \sqrt{21}}{192}$ &$\frac{5 \sqrt{42}}{192}$ & $\frac{\sqrt{21}}{192}$& $-\frac{5 \sqrt{7}}{64}$\\
$\frac{7 \sqrt{6}}{64}$& $0$ &$- \frac{5}{32}$ &$- \frac{3 \sqrt{42}}{64}$ &$- \frac{5 \sqrt{21}}{64}$ &$\frac{5 \sqrt{7}}{64}$ &$\frac{\sqrt{14}}{64}$ &$\frac{5 \sqrt{7}}{64}$ & $-\frac{5 \sqrt{21}}{64}$\\
$- \frac{35 \sqrt{6}}{192}$&$- \frac{5 \sqrt{3}}{96}$ & $\frac{3}{16}$&$- \frac{5 \sqrt{42}}{192}$ &$\frac{5 \sqrt{21}}{192}$ &$- \frac{3 \sqrt{7}}{64}$ &$\frac{5 \sqrt{14}}{64}$ & $-\frac{5 \sqrt{7}}{64}$&$-\frac{3 \sqrt{21}}{64}$ \\
$- \frac{\sqrt{10}}{64}$& $- \frac{7 \sqrt{5}}{32}$& $\frac{\sqrt{15}}{16}$&$\frac{\sqrt{70}}{64}$ & $-\frac{ \sqrt{35}}{64}$&$-\frac{\sqrt{105}}{64}$ &$-\frac{\sqrt{210}}{64}$ &$\frac{\sqrt{105}}{64}$ & $-\frac{3 \sqrt{35}}{64}$\\
$\frac{7 \sqrt{10}}{64}$& $- \frac{\sqrt{5}}{16}$& $-\frac{\sqrt{15}}{32}$&$- \frac{3 \sqrt{70}}{64}$ &$\frac{3 \sqrt{35}}{64}$ &$- \frac{\sqrt{105}}{64}$ &$\frac{\sqrt{210}}{192}$ &$-\frac{7 \sqrt{105}}{192}$ &$-\frac{\sqrt{35}}{64}$ \\
$- \frac{\sqrt{30}}{64}$&$\frac{\sqrt{15}}{16}$ & $- \frac{3 \sqrt{5}}{32}$&$\frac{\sqrt{210}}{64}$ &$- \frac{ \sqrt{105}}{64}$ &$-\frac{3\sqrt{35}}{64}$ &$-\frac{3 \sqrt{70}}{64}$ &$-\frac{3 \sqrt{35}}{64}$ & $-\frac{\sqrt{105}}{64}$\\
\bottomrule
\end{tabular}
\end{center}
\caption{\small{Island subduction coefficients of $[4,3,2,1] \downarrow [3,2,1] \otimes [3,1]$ for each multiplicity copy. The coefficients are listed in the lexicographic ordering (when they are read from left to right and top to bottom) and they have the same $m$ along the rows and $m_2$ down the coloumns.}}
\end{table}
The first multiplicity-three case for the subduction problem in symmetric groups accours in $[4,3,2,1] \downarrow [3,2,1] \otimes [3,1]$ of $S_{10} \downarrow S_6 \times S_4 $. From the hook rule~\cite{Fulton}, the representation $[4,3,2,1]$ has dimension $f^{\lambda}=768$, $[3,2,1]$ has dimension $f^{\lambda_1}=16$ and $[3,1]$ dimension $f^{\lambda_2}=3$. Thus we have $f^{\lambda} f^{\lambda_1} f^{\lambda_2}=36864$ SDCs to evaluate. Many of such coefficients are zero via the selection rule provided in the previous section. Now, the number of islands is given by the dimension of $[3,2,1]$, i.e. $16$. But, from theorem $2$, we only need to determine the SDCs corresponding to {\it one} island. Because the number of standard skew-tableaux of shape $[4,3,2,1]/[3,2,1]$ is $f^{\lambda/\lambda_1}=24$, the island is composed of $f^{\lambda/\lambda_1} f^{\lambda_2} =72$ nodes which correspond to our unknowns.  
\par 
We organize the nodes by the lexicographic ordering: first we order the tableaux and then the triplet which forms each node. We choose the usual Yamanouchi convention~\cite{Chenbook} to fix the phase freedom: we impose the {\it first non-zero} SDC to be {\it positive}.
\par 
From subduction graph~\cite{Chilla} and by using a suitable Mathematica program~\cite{math}, we generate the homogeneous linear sistem required to obtain the SDCs. Then we find the kernel of the subduction matrix which provides a non-orthonormalized form for the coefficients. The solution space has dimension $3$ (multiplicity). We orthonormalize the SDCs in such a way that the conditions (\ref{orton1}) and (\ref{orton2}) hold.
\par 
In table $1$ we deal with the three copies for the SDCs (with multiplicity labels $1$, $2$ and $3$, respectively). Such coefficients are listed in the lexicographic ordering (when they are read from left to right and up to down) and they have a fixed $m$ along the rows and $m_2$ down the coloumns. Multiplicity separation can be choosen in such a way that the coefficients are expressed as a single surd of the form $a\sqrt{b}/c$, with $a$, $b$ and $c$ integers.  
\par 
By coniugation of $[4,3,2,1] \downarrow [3,2,1] \otimes [3,1]$, i.e. $[4,3,2,1] \downarrow [3,2,1] \otimes [2,1,1]$, we have another multiplicity three case for the subduction problem. The new SDCs are related to the previous ones. Denoted as $\tilde{m}$ the skew-tableau conjugate to $m$ and as $\tilde{m}_1$ and $\tilde{m}_2$ the tableaux conjugate to $m_1$ and $m_2$ respectively, we have, for the $m_1$-island, the following symmetry conditions
\begin{equation}
\langle \tilde{\lambda}; \tilde{m}|\tilde{\lambda}_1, \tilde{\lambda}_1; \tilde{m}_1, \tilde{m}_2\rangle_1 = \Lambda^{[4,3,2,1]/[3,2,1]}_{m/m_1} \ \Lambda^{[3,1]}_{m_2} \ \langle \lambda;m| \lambda_1, \lambda_2 ;m_1,m_2 \rangle_3 
\end{equation}
\begin{equation}
\langle \tilde{\lambda}; \tilde{m}|\tilde{\lambda}_1, \tilde{\lambda}_2; \tilde{m}_1, \tilde{m}_2\rangle_2 = \Lambda^{[4,3,2,1]/[3,2,1]}_{m/m_1} \ \Lambda^{[3,1]}_{m_2} \ \langle \lambda;m| \lambda_1, \lambda_2 ;m_1,m_2 \rangle_2 
\end{equation}
\begin{equation}
\langle \tilde{\lambda}; \tilde{m}|\tilde{\lambda}_1, \tilde{\lambda}_2; \tilde{m}_1, \tilde{m}_2\rangle_3 = \Lambda^{[4,3,2,1]/[3,2,1]}_{m/m_1} \ \langle \lambda;m| \lambda_1, \lambda_2 ;m_1,m_2 \rangle_1, 
\end{equation}
where $\Lambda^{\lambda}_{m}$ are the phase factors of the Yamanouchi basis~\cite{Butler} for the irrep $[\lambda]$ and 
\begin{equation}
\Lambda^{\lambda/\lambda_1}_{m/m_1} =\Lambda^{\lambda}_m \Lambda^{\lambda_1}_{m_1} \ \ \ \ \ \text{m compatible with $m_1$}
\nonumber 
\end{equation}
(notice that, if $m$ and $m_1$ are compatible, $m/m_1$ represents the skew-tableau of shape $\lambda/\lambda_1$ obtained by removing the first $n_1$ boxes from $m$). Denoted as $m'$ and $m_2'$ the ordering number (by lexicografic ordering) for $m/m_1$ and $m_2$ respectively, our Mathematica computation provides
\begin{equation}
\begin{array}{cc}
\Lambda^{[4,3,2,1]/[3,2,1]}_{m'} = - (-1)^{\frac{m'(m'-1)}{2}} & \text{with $m' \in \{1,2,3, \dots, 24 \}$} \\
\Lambda^{[3,1]}_{m_2'} = - (-1)^{m_2'} & \text{with $m_2' \in \{ 1,2,3 \}$}.
\end{array}
\end{equation}
\section{Summary}
We have considered transformations between split bases and standard bases of the symmetric group $S_n$. A selection rule which allows to determine the vanishing SDCs and to organize the other ones in blocks (named islands) was given. We have proven that all islands produce the same values for the SDCs and thus only a very smaller number of them really needs to be evaluated. The linear equation method, described in terms of a reduced subduction graph, provides a systematic and optimizated tool to calculate the unknown transformation coefficients. 
\par  
As a significative example, the first multiplicity-three cases, $[4,3,2,1] \downarrow [3,2,1] \otimes [3,1]$ and its conjugate one $[4,3,2,1] \downarrow [3,2,1] \otimes [2,1,1]$ for $S_{10} \downarrow S_6 \times S_4$, were dealt in detail: we have given the suitable orthonormalized transformation coefficients relative to each multiplicity copy descending from the Yamanouchi phase convention.
\par 
We have implemented a Mathematica code which provides the solution for every multiplicity cases and we have obtained the results up to the decomposition of $[5,4,3,2,1]$ of $S_{15}$ into $[4,3,2,1] \otimes [4,1]$ of $S_{10} \times S_{5}$ which is the first multiplicity-four example. In this case $f^{[5,4,3,2,1]}= 292864$, $f^{[4,3,2,1]}=768$ and $f^{[4,1]}=4$, therefore the SDCs are a total of $ f^{[5,4,3,2,1]}f^{[4,3,2,1]} f^{[4,1]} = 899678208$. Because $f^{[5,4,3,2,1]/[4,3,2,1]}=120$, the number of SDCs relative to an island only is $ f^{[5,4,3,2,1]/[4,3,2,1]} f^{[4,1]} = 480$. Interested readers may contact the author for further information. 
\par
In table 2 we deal with some subduction cases, the relative multiplicity, the number of unknowns involved in the primal linear equation system and the effective number of needed SDCs, after the application of the selection and identity rules. It is evident the drastic reduction of the number of unknowns for the subduction problem.   
\begin{table}
\label{tabmult}
\begin{center}
\begin{tabular}{cccc}
\toprule
$[\lambda]\downarrow [\lambda_1]\otimes [\lambda_2]$ & $\{\lambda; \lambda_1, \lambda_2 \}$ & $f^{\lambda} f^{\lambda_1} f^{\lambda_2}$& $f^{\lambda/\lambda_1} f^{\lambda_2}$ \\
\midrule
$[4,2]\downarrow [2,1]\otimes [2,1]$ & $1$ & $36$ & $6$ \\
$[3,2,1]\downarrow [2,1]\otimes [2,1]$ & $2$ & $64$  & $12$  \\
$[4,2,1]\downarrow [3,1]\otimes [2,1] $ & $2$ & $210$ & $12$ \\
$[4,3,2]\downarrow [3,2]\otimes [3,1]$ & $2$ & $2520$ & $36$ \\
$[4,3,2,1]\downarrow [3,2,1]\otimes [3,1]$ & $3$ & $36864$ & $72$ \\
$[5,4,3,2]\downarrow [4,3,2]\otimes [3,2]$ & $3$ & $40360320$ & $300$ \\
$[5,4,3,2,1]\downarrow [4,3,2,1]\otimes [4,1]$ & $4$ & $899678208$ & $480$ \\
$[6,5,4,3,2,1]\downarrow [5,4,3,2,1]\otimes [5,1]$ & $5$ & $1611839486033920$ & $3600$ \\
\bottomrule
\end{tabular}
\end{center}
\caption{\small{Some examples of subduction with the relative multiplicity, the primal number of involved SDCs and the island dimension.}}
\end{table}
Thus, by a reduced version of the subduction graph, we obtain an improved approach to {\it high} dimension subduction problem in symmetric groups and the collateral representation theory which are often useful in many-body calculations for quantum physics and in nuclear and high energy physics issues.
\ack
The author thanks Massimo Campostrini and Matteo Ruggiero for fruitful discussions. The computing resources of ``Collegio Timpano'' (Scuola Normale Superiore - Pisa, Italy) are gratefully acknowledged.  

 \end{document}